\documentclass[aps,prd,twocolumn,groupedaddress]{revtex4-1}
\usepackage{amssymb,amsmath}
\usepackage{graphicx}
\usepackage{subfigure}
\usepackage{hyperref}
\hypersetup{
colorlinks=true,
linkcolor=red,
citecolor=blue,
filecolor=cyan,
urlcolor=magenta,
}
\begin{document}

\title{Probing the Sun's inner core using solar neutrinos: a new diagnostic method}
\author{Il\'idio Lopes}
\email[]{ilidio.lopes@ist.utl.pt;ilopes@uevora.pt}
\affiliation{CENTRA, Instituto Superior T\'ecnico, 1049-001 Lisboa, Portugal}
\affiliation{}
\affiliation{Departamento de F\'isica, Universidade de \'Evora,  7002-554 \'Evora, Portugal}

\date[Published in Phys. Rev. D 88, 045006, ]{7 August 2013}

\begin{abstract}
 The electronic density in the Sun's inner core is inferred from the $^8B$, $^7Be$ and $pep$ neutrino 
 flux measurements of the Super-Kamiokande, SNO and Borexino experiments.
 We have developed a new method in which we use the KamLAND detector determinations 
 of the neutrino fundamental oscillation parameters: the mass difference and the vacuum oscillation angle.
 Our results suggest that the solar electronic density in the Sun's inner core (for a radius 
 smaller than 10\% of the solar radius) is well above the current prediction of the standard  solar model, and  
 by as much as 25\%.  A potential confirmation of these preliminary findings can be achieved 
 when neutrino  detectors are able to reduce the error of the electron-neutrino survival probability by a factor of 15. 
\end{abstract}

\maketitle

\section{Introduction\label{sec-intro}}

Traveling close to the speed of light and with a very large mean free path, 
solar neutrinos provide a unique and powerful tool to study the properties of the plasma in the Sun's core.
During the last two decades, the increase in the number of neutrino detectors,
accompanied  by a significant improvement in the accuracy of the measurements 
has led to major progress in the research field of neutrino physics.

Among many breakthroughs in solar neutrino physics, two are particularly worth mentioning- due to their relevance for this work.
In 2001, the Sudbury Neutrino Observatory (SNO) experiment~\cite{2001PhRvL..87g1301A}, following in the 
footsteps of previous neutrinos experiments \cite[e.g.][]{2011RPPh...74h6901T}, definitely confirmed the  
neutrino flavor oscillations in vacuum and matter. 
This theoretical model was first suggested by Pontecorvo~\cite{1958JETP....6..429P} to explain oscillations in vacuum, 
and later extended by~Wolfenstein~\cite{1978PhRvD..17.2369W}, Mikheyev
and Smirnov~\cite{1986NCimC...9...17M}  
to include the oscillations of neutrinos in matter.
This is the reason why this last oscillation mechanism is also called the 
Mikheyev-Smirnov-Wolfenstein (MSW) effect.

In the following year another important discovery occurred: the Kamiokande 
Liquid Scintillator Anti-Neutrino Detector (KamLAND)~\cite[][]{2003PhRvL..90b1802E} 
measured the flux of antineutrinos 
from distant reactors confirming the oscillation nature of neutrinos.  Furthermore, KamLAND found 
that a unique combination of oscillation parameters could explain the neutrino oscillations data, 
the so-called large mixing angle solution.  More significantly, for the first time it was possible to measure 
the parameters related with the flavor oscillations of neutrinos in vacuum, 
from a source of neutrinos that is not the Sun. 
The three-flavor neutrino oscillations based only on KamLAND data
analysis~\cite{2011PhRvD..83e2002G,2008PhRvL.100v1803A} 
give $\Delta m^2_{21} = 7.49\pm 0.20\times 10^{-5}\; eV2$ 
and $\tan^2{\theta_{21}} = 0.436\pm 0.102 $. 
The two-flavor neutrino oscillations  data analysis ($\theta_{13}=0$) 
gives identical results, but the value of $\tan^2{\theta_{21}}$ increases by 11\%.

Once the fundamental parameters of neutrino flavor oscillations are determined 
independently of the neutrinos coming from the Sun (at least in the case of the KamLAND experiment), 
it is reasonable to use, or at least to discuss, the possibility of using solar neutrino flux 
measurements to probe the Sun's interior. Although our understanding of the basic physical
mechanisms occurring inside the Sun is quite robust, there is still a certain number of 
unknown processes in the solar core that neutrinos could help to resolve~\cite[see Ref. ][ and references therein]{2012RAA....12.1107T}.  Furthermore, 
if we wish to use  the Sun as a cosmological
tool~\cite[e.g., Refs. ][]{2012ApJ...752..129L,2010Sci...330..462L,2010ApJ...722L..95L,2010PhRvD..82h3509T,2010PhRvD..82j3503C,2002MNRAS.331..361L,2002PhRvL..88o1303L}
in an identical manner to how neutron stars~\cite[e.g., Refs. ][]{2012PhRvL.108s1301K,2011PhRvD..83h3512K,2011PhRvL.107i1301K,2010PhRvD..82f3531K,2012arXiv1212.4075K}
are used to constrain the dark matter particle properties, 
then this will only be possible if we have reliable methods to diagnose the solar core.

Most of the progress achieved in describing accurately the physical processes occurring in the Sun's 
interior is owed to helioseismology~\cite[e.g., Ref. ][]{2011RPPh...74h6901T}: 
the present seismic data allows the determination of the sound-speed and 
density radial density profiles to be obtained with high accuracy, between the solar surface and the first layers of the  nuclear region~\cite{2009ApJ...699.1403B,1997SoPh..175..247T}.
Comparing the present Sun's structure as predicted by the standard solar 
model (SSM) with the helioseismological data, 
the sound-speed difference is at most of the order of 2\% (mainly in the radiative region) 
and the density difference is of the order of 10\%  (in the convective zone).
The origin of the sound-speed difference is unknown, probably related to some physical 
mechanisms occurring during the evolution of the star~\cite{2012RAA....12.1107T}, 
but the  density difference has been  linked to a poor description of 
the solar convection~\cite{1997ApJ...480..395A,2009LRSP....6....2N}.
 
In the last  few years, several theoretical models have been put forward to explain this  
discrepancy between theory and helioseismological data. Most of these proposals are 
able to reduce the sound-speed difference. Among the various proposals,  
we choose to mention three of them, which are also validated by stellar observations.  
This is the case with solar models with no standard   pre-main-sequence evolution, 
like solar evolution models that take into account  the temporal evolution 
of the solar internal rotation~\cite{2010ApJ...715.1539T}, and solar models for which during a 
certain period of their  pre-main-sequence evolution, 
the star has its mass changed by a mechanism of mass-loss
or accretion~\cite{2010ApJ...713.1108G,2011ApJ...743...24S}. 
  
Another physical process that has been recently considered important for 
Sun-like stars with planetary systems, is the possibility that the star 
during its formation or even in its pre-main-sequence phase, could have 
its internal metallicity increased due to  the migration of heavy nuclei 
from the planetary disk towards the core of the protostar~\cite{Lopes:2013ux}.
This scenario has been suggested by recent high-precision spectroscopic observations of solar 
twins with and without planetary systems~\cite{2009ApJ...704L..66M,2009A&A...508L..17R}.  
In particular, it was found that the former group of stars presents a larger amount of metals 
in the surface than the latter group. Up to now, how exactly this mechanism  operates is still unknown, 
but it has been shown to be linked to the formation of Earth-like planets.
If this process occurs, this could increase significantly the abundances of elements such as  
carbon, nitrogen and oxygen, increasing the overall metallicity of the stellar interior.  
Therefore, the amount of heavy elements   
like carbon, nitrogen and oxygen in the Sun's interior 
can be well above the values presently measured in 
the solar photosphere~\cite{2009ARA&A..47..481A}.  
The observations suggest that stars like the Sun could have a 30\% excess of metals 
in their radiative interior when compared with current values predicted by the 
standard solar model~\cite{2011RPPh...74h6901T,2012RAA....12.1107T}.
Furthermore, the sound-speed difference is reduced to a value qualitatively 
close to the sound-speed difference obtained  with the previous metal 
abundance measurements~\cite{1998SSRv...85..161G}.  
This issue is particularly relevant for the neutrinos produced in the 
nuclear reactions of the carbon-nitrogen-oxygen (CNO) cycle. 
If the excess in abundances occurs for Carbon, Nitrogen and Oxygen elements,
the neutrino fluxes produced in the CNO cycle 
will be well above the values predicted by the standard solar model~\cite{Lopes:2013ux}. 

As shown, several physical mechanisms are concurrent with each other towards reducing the sound-speed difference 
in the solar interior, but the current solar data (including helioseismology data) 
do  not allow us  unequivocally to determine which are the best  proposals.
A strategy going forward to resolve this issue is to investigate new techniques 
to diagnose the plasma of the solar interior including the solar inner core.
In particular, this will be done by constraining quantities other than the sound-speed profile, 
such as the electronic density or the plasma density.

Presently, due to the low amount of low-degree acoustic modes, 
there is still a large uncertainty in the inversion of the sound-speed 
profile in the solar inner core, and consequently, 
the seismic diagnostic of this region is quite inaccurate.
Even  if it is possible to have some information about the sound-speed in this region, 
our knowledge about the local density is much more uncertain.
Naturally, the most reliable hope to probe the density in the Sun's core is to use gravity modes, 
although their existence  must still be confirmed ~\cite{2004ApJ...604..455T}. 
The solar neutrinos are a natural alternative for diagnosing the solar interior,
provided that the precision in the measurement of neutrino fluxes is obtained with the required accuracy. 
If this experimental goal is succeeded,  
it will provide a major contribution for resolving this problem.  

A first attempt to obtain the electronic density in the Sun's 
interior was made by Balantekin {\it et al.}~\cite{1998PhLB..427..317B}. In their work, the authors computed 
the electronic density as an expansion in powers of the local density plasma, under the 
hypothesis that neutrino oscillations occur with a small-angle MSW solution.
The result obtained allows  a relatively good analytical representation of the electronic density in most of the radiative region.
The method becomes imprecise only in the core of the Sun below 20\% of the solar radius.

Finally, it is worth mentioning that there is still some uncertainty on the basic parameters of neutrino oscillations, 
which in turn can introduce some uncertainty in the total fluxes of neutrinos of different flavors. 
However, as   was pointed out by Bhat {\it et al.}~\cite{1998PhRvL..81.5056B}, this effect 
is much smaller than the MSW effect discussed in this work. In particular,Balantekin and Malkus ~\cite{2012PhRvD..85a3010B} have studied the impact of the 
third  neutrino oscillation mixing angle in vacuum  on the  
neutrino flavor oscillations related with matter, and have found that the impact was negligible.
A review of the properties of the propagation of neutrinos in matter can be found in Balantekin~\cite{1999PhR...315..123B}.
 
In this paper, we focus on the inner nuclear region of the Sun ($r\le 0.1 R_\odot$), and we will present and discuss a new diagnostic method that constrains the electronic density that comes from current neutrino flux measurements in this region. The method is based in determining the electronic density in the Sun's inner core, as a linear correction to the present values of solar neutrino fluxes obtained for the present standard solar model.
 What makes this new diagnostic particularly appealing is the possibility of constraining the electronic density at a distance of 5\% from the center of the Sun, a very central region which is very difficult to probe by any  other methods.

  
\section{The Solar Neutrino Energy Spectrum}
Neutrinos are produced in the Sun's interior through several reactions of the nuclear network,
 usually known as proton-proton ($PP$) chains and  carbon-nitrogen-oxygen (CNO) cycle.
Figure~\ref{fig1} shows the neutrino spectra of our standard solar model~\cite{2013ApJ...765...14L}  
computed with an  updated version of the stellar evolution code {\sc cesam}~\cite{1997A&AS..124..597M}. 
This version of {\sc cesam} has an up-to-date  and very refined microscopic physics
(equation of state, opacities, nuclear reaction  rates,
and an accurate treatment of microscopic diffusion of heavy elements), 
including the solar mixture of Asplund {\it et al.}~\cite{2009ARA&A..47..481A}  
and nuclear reaction rates from NACRE Compilation~\cite{2011RvMP...83..195A,1999NuPhA.656....3A}.
The solar models are calibrated to the present solar radius 
$R_\odot= 6.9599 \times 10^{10} \;{\rm cm}$,  luminosity  $L_\odot = 3.846 \times 10^{33} \; {\rm erg\; s}^{-1}$, 
mass $M_\odot = 1.989 \times 10^{33} \;{\rm g}$, and 
age $t_\odot = 4.54\pm 0.04\; {\rm Gyr}$~\cite[e.g. Ref. ][]{2011RPPh...74h6901T}. 
The models are required to have a fixed value of the photospheric ratio of metal abundance over hydrogen abundance 
in agreement with the used solar mixture. The total neutrino fluxes (on Earth) predicted by this model are the following:
$\phi(pep)=1.4\; 10^{8} $, $\phi(^7Be)=4.7\; 10^{9}$, $\phi(^8B)=5.3\; 10^{6}$,  
$\phi(^{13}N)=5.3\; 10^{8}$, $\phi(^{15}O)=4.5\; 10^{8}$,  $\phi(^{17}F)=5.0\; 10^{6}$ and  
$\phi(pp)=5.9\; 10^{10} $, in units of ${ \rm cm^{-2}\;s^{-1} }$.
This solar model is in agreement with the most current helioseismology data and is identical to others published in the
literature~\cite[e.g., Refs. ][]{2011ApJ...743...24S,2010ApJ...715.1539T,2010ApJ...713.1108G,2011ApJ...743...24S,2009ApJ...705L.123S,2010ApJ...715.1539T}. The neutrino fluxes of $^8B$, $^7Be$ and $pep$ are in agreement with the current solar neutrino experiments.
The new NACRE table~\cite{2011RvMP...83..195A} presents a set of S factors   
slightly different of previous ones~\cite{1999NuPhA.656....3A}, 
but with a smaller error bar.
Therefore, the $^8B$, $^7Be$ and $pep$ total neutrino fluxes are almost the same ones 
found on previous predictions, the differences are mainly due to the change 
of S factors of few nuclear reactions like $^3He(^3He,2p)^4He$ 
and $^7Be(p,\gamma)^8B$.
Nevertheless, the standard solar model computed with  this 
new set of laboratory cross-section measurements~\cite{2011RvMP...83..195A}  
predicts solar neutrino fluxes  in excellent 
agreement with experimental data~\cite{2013PhRvD..87d3001S}. 
Moreover, the impact of this update of the NACRE table on the Sun's core electronic 
density is negligible,  because the ratios of neutrino flavor fluxes 
are almost independent of the total neutrino flux for each neutrino source.

Figure~\ref{fig2} shows the different neutrino emission sources inside the Sun. 
The nuclear reactions are ordered inside the Sun according to the temperature 
required for the fusion reaction between reacting nuclei to occur.
Naturally, the nuclear reactions between heavy nuclei are located nearer the center, 
than reactions between lighter nuclei.  In Fig.~\ref{fig2} we show the electron-neutrino source function $ \Phi (r)$
related to the electron-neutrinos produced in the $PP$ chain 
($ pp$, $pep$, $^8B$ and $^7Be$)  and  CNO  cycle ($^{13}N$, $^{15}O$, $^{17}F$) nuclear reactions.

The neutrino sources produced by the nuclear reactions of the $PP$ chain are located between 
the center and 30\% of the solar radius. The $pp$ neutrino source extends from the   
center up to 30\% of the solar radius 
with its maximum located at 10\% of the solar radius. 
The other $PP$ chain neutrino sources are   $pep$,  $^7Be$ and $^8B$ and have  emission shell with  widths of  
22\%, 18\% and 10\% respectively, and  maximums occurring at 8.6\%,  5.8\% and 4.5\% of the solar radius, respectively.
In particular, it is worth noticing that the $pp$ and $pep$  nuclear reactions are strongly 
dependent  on the total luminosity of the star.  
Similarly, the neutrino sources related with the  CNO cycle of  nuclear reactions (cf. Fig.~\ref{fig2}), 
$^{15}O$, $^{17}F$ and  $^{13}N$ neutrinos occur in emission shells with a width of 16\% of the solar radius. 
The maximum neutrino production of these sources occurs at 5\% of the solar radius.
$^{13}N$ neutrinos have a second emission shell located between 12\% and 25\% of the Sun's radius with its
maximum occurring at 16\% of the solar radius.

It is important to note that an increase or decrease of the temperature in the core of the Sun (caused by the presence of some known or unknown physical process) does not change much the location of  the neutrino emission region $\Phi(r)$, even if the total  neutrino fluxes change significantly (cf. Fig~2). This is due to the strong dependence of neutrino  nuclear reactions on the temperature. 

In the case of solar models for which the central temperature is changed  by a few percent,
the  neutrino emission regions
(i.e., the $^8B$, $^7Be$ and $pep$  neutrino emission regions among others) does not change  much,  although the total neutrino fluxes are strongly affected. The position of the maximum of $\Phi(r)$ of the different neutrino sources changes by less than 1\%.

\section{Neutrino Flavor Oscillations}

The theory of neutrino flavor oscillations~\cite[i.e., Ref. ][]{2010LNP...817.....B}
describes the propagation of neutrinos in space~\cite[][]{1958JETP....6..429P} and the interaction 
of neutrinos with matter~\cite[][]{1986NCimC...9...17M,1978PhRvD..17.2369W}. 
Yuksel~\cite{2003PhRvD..68k3002B} showed that the survival probability of solar neutrinos calculated 
in a model with two neutrino flavor oscillations or three neutrino flavor oscillations  
have very close values.

In the Sun, the electron-neutrino survival or appearance 
probabilities depend only on three fundamental oscillation parameters: the mass difference  
$\Delta m_{12}$ and the angles $\theta_{21}$  and $\theta_{13}$.  In the absence of the MSW effect caused by the 
Earth's globe, the survival probability during the day of a three-flavor neutrino oscillation 
can be reduced to a modified two-flavor neutrino oscillation model that is accurately described as
\begin{eqnarray} 
P_{\nu_e}(E_\nu, r) = \cos^4{\theta_{13}}\;P_{2\nu_e}(E_\nu, r)+\sin^4{\theta_{13}},
\label{eq-A}
\end{eqnarray} 
where  $ E_\nu $ is the energy of the neutrino and 
$P_{2\nu_e}(E_\nu, r) $ is the probability in the case of a two-flavor oscillation  model~\cite[e.g., Refs. ][]{2011PhRvD..83e2002G,2010LNP...817.....B,1932ZPhy...78..847L}. 
$P_{2\nu_e}(E_\nu, r) $  is given by
\begin {eqnarray} 
P_{2\nu_e}(E_\nu, r) =  \frac {1} {2} + \frac {1} {2}  \cos {(2 \theta_{21})} \cos{(2 \theta_m)}
\label{eq-B}
\end {eqnarray} 
where $\theta_{m}$ is the mixing angle at the production source inside the Sun.
In the derivation of this formula  it was assumed  that the first-order correction to  
the propagation of neutrinos in matter is valid  once the adiabatic condition  
is verified in the radiative region of the solar interior~\cite{1932ZPhy...78..847L,1932RSPSA.137..696Z}. 
The phase $ \theta_m $ is the most important term in this analysis, 
since it depends on the electron density of the Sun's core. The mixing angle $\theta_m$ is given by 
\begin{eqnarray} 
\sin{(2\theta_m)}= 
\frac{\sin{(2\theta_{12})}} 
{\sqrt{(V_m-\cos{(2\theta_{12})})^2+ \sin^2{(2\theta_{12})}} } ,
\label{eq-C}  
\end{eqnarray}
where $V_m$  is a function of solar radius.
The values of $ \Delta m_{12}^2 $  and $ \theta_{21} $ can be  solely determined 
from  neutrino experiments  (not using the neutrinos coming from the Sun),
such as the  KamLAND reactor experiment~\cite{2003PhRvL..90b1802E}.
$V_{m} $  is a function of solar plasma density,  given by
\begin{eqnarray} 
V_m (E_\nu,r) = 2 \sqrt{2} G_f \; n_e (r)\; E_\nu\;\cos^2{(\theta_{13})}/\Delta m_{21},
\label{eq-D}  
\end{eqnarray}
where $ G_f $ is the Fermi constant and $ n_e (r) $ is the electron density of plasma. 
The  electron density $ n_e (r) =N_{o}\; \rho(r)/\mu_e(r) $, 
 where $\mu_e(r)$ is the mean molecular weight per electron, $\rho(r)$ the density of matter
 in the solar interior and $N_{o}$ the Avogadro's number.

The radial neutrino emission profile of electron neutrinos 
is specific for each nuclear reaction (cf. Fig.~\ref{fig2}).  It follows that the average survival probability of electron neutrinos 
for each of the emission neutrino sources,  $\langle P_{\nu_e } (E_\nu)\rangle$  is given by
\begin{eqnarray} 
\langle P_{\nu_e } (E_\nu)\rangle = 
A^{-1} \int P_{\nu_e} (E_\nu,r)\Phi (r) 4\pi \rho(r) r^2 dr 
\label{eq-F}
\end{eqnarray}  
where $ A=  \int \Phi (r) 4 \pi \rho (r) r^ 2  \;dr $ is a normalization constant.
In the following $\langle \cdots\rangle $ defines the
$ \Phi (r) $ weight averaged of a certain quantity, as defined in the previous equation.

The electron-neutrino survival probability functions $\langle P_{\nu_e } (E_\nu)\rangle$ for different nuclear
reactions are shown in Fig.~\ref{fig3}. 
The  survival probabilities of electron-neutrinos were computed by using
the fundamental parameters of solar neutrino oscillations in vacuum
namely $\Delta m_{12}$ and $\theta_{12}$,  as determined by the KamLAND experiment~\cite{2011PhRvD..83e2002G}.
Although the contribution related to $\theta_{13}$ is minor, we take its contribution into 
account by choosing $\theta_{13}=9 {\rm deg}$, a value that is in agreement with current experimental measurements, 
$8.96^{+0.45}_{-0.51} \deg $~\cite{2012PhRvD..86a3012F}  and $ 9.06^{+0.50}_{-0.57} \deg $~\cite{2012PhRvD..86g3012F}. Furthermore,  Balantekin and Malkus~\cite{2012PhRvD..85a3010B}  have shown that
the electron-neutrino survival probability functions are not very sensitive to the mixing $\theta_{13}$  
for the neutrino energy range where the MSW effect is dominant, 
at least in the energy range where the neutrino flux measurements are taken.

\begin{figure}
\centering
\includegraphics[scale=0.5]{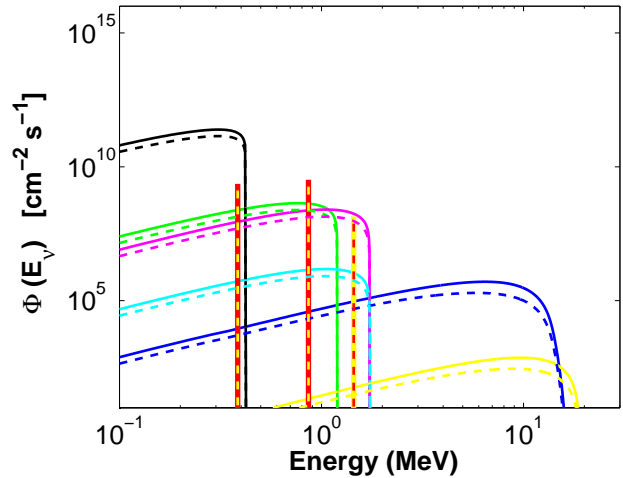}
\caption{The solar neutrino energy spectrum predicted by our standard solar model. 
The solid curves correspond to the  total (electron-flavor)  neutrino fluxes produced in the various nuclear reactions of the  
proton-proton chains  and carbon-nitrogen-oxygen  cycle. The dashed curves correspond to  electron-neutrino fluxes of the various nuclear reactions after neutrino flavor conversion.  The color curves define the following neutrino sources 
from the proton-proton chain reactions: $^8B$ (blue curve),  $^7Be$ (two red-yellow lines), $ pep$ (yellow-red line), 
$hep$ (yellow curve) and $pp$ (black curve); and the following neutrino sources from the 
carbon-nitrogen-oxygen: $^{13}N$ (green curve), $^{15}O$ (magenta curve) and  $^{17}F$ (cyan curve). 
The neutrino fluxes from continuum nuclear sources are given in units
of $\rm cm^{-2}s^{-1}Mev^{-1}$. The line neutrino fluxes are given in $\rm cm^{-2}s^{-1}$.
}\label{fig1}
\end{figure}

\begin{figure}
\centering
\includegraphics[scale=0.5]{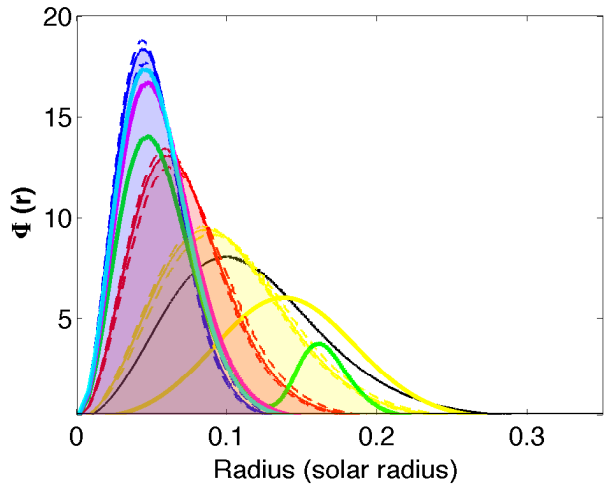}
\caption{ 
The electron-neutrino fluxes produced in the various nuclear reactions.
The $\Phi (r)$ for which neutrino fluxes have been measured experimentally 
($^8B$, $^7Be$ and $pep$) are indicated with a shaded area. 
 If the central temperature 
of the standard solar model changes by $+2.5\%$  ($-4\%$) and density changes 
accordingly ($+7.4\%$ and $-12.7 \%$) to keep the energetic balance,
the neutrino fluxes change by $+70\%$ ($-60\%$) for $^8B$, by $ 30\%$ ($-38\%$) 
for the $^7Be$ neutrino emission lines and by  $13\%$ ($-20\%$) for the $pep$  emission line. 
The  function $\Phi(r)$ for these two cases is indicated by dashed curves.
The color scheme is the same as for Fig.~\ref{fig1}. In particular,  $^7Be$ (two red-yellow lines)
and $ pep$ (yellow-red line) in  Fig.~\ref{fig1} correspond this figure's red and  yellow shaded areas,
respectively.
}
\label{fig2}
\end{figure}

\begin{figure}
\centering{
\includegraphics[scale=0.5]{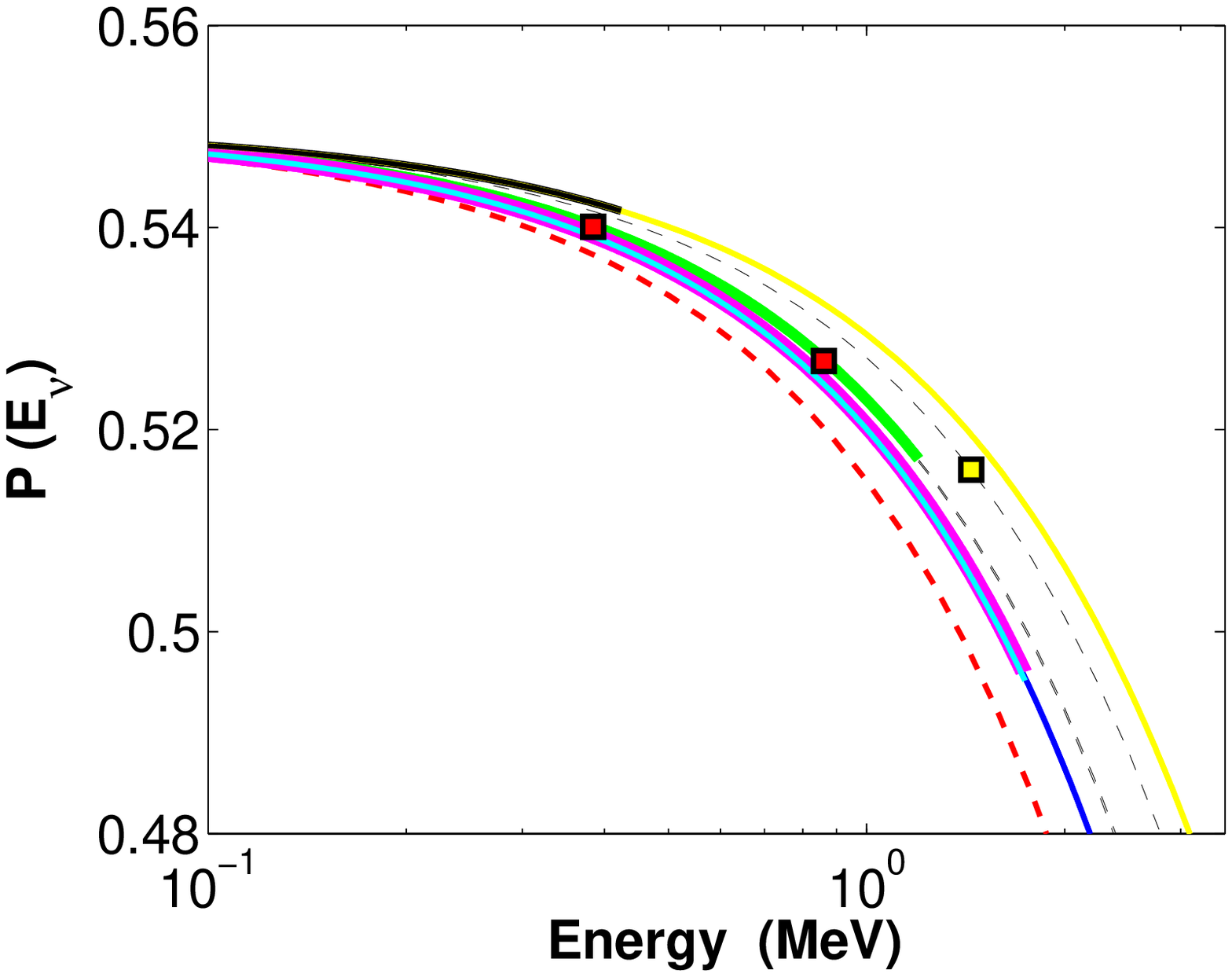}
\includegraphics[scale=0.5]{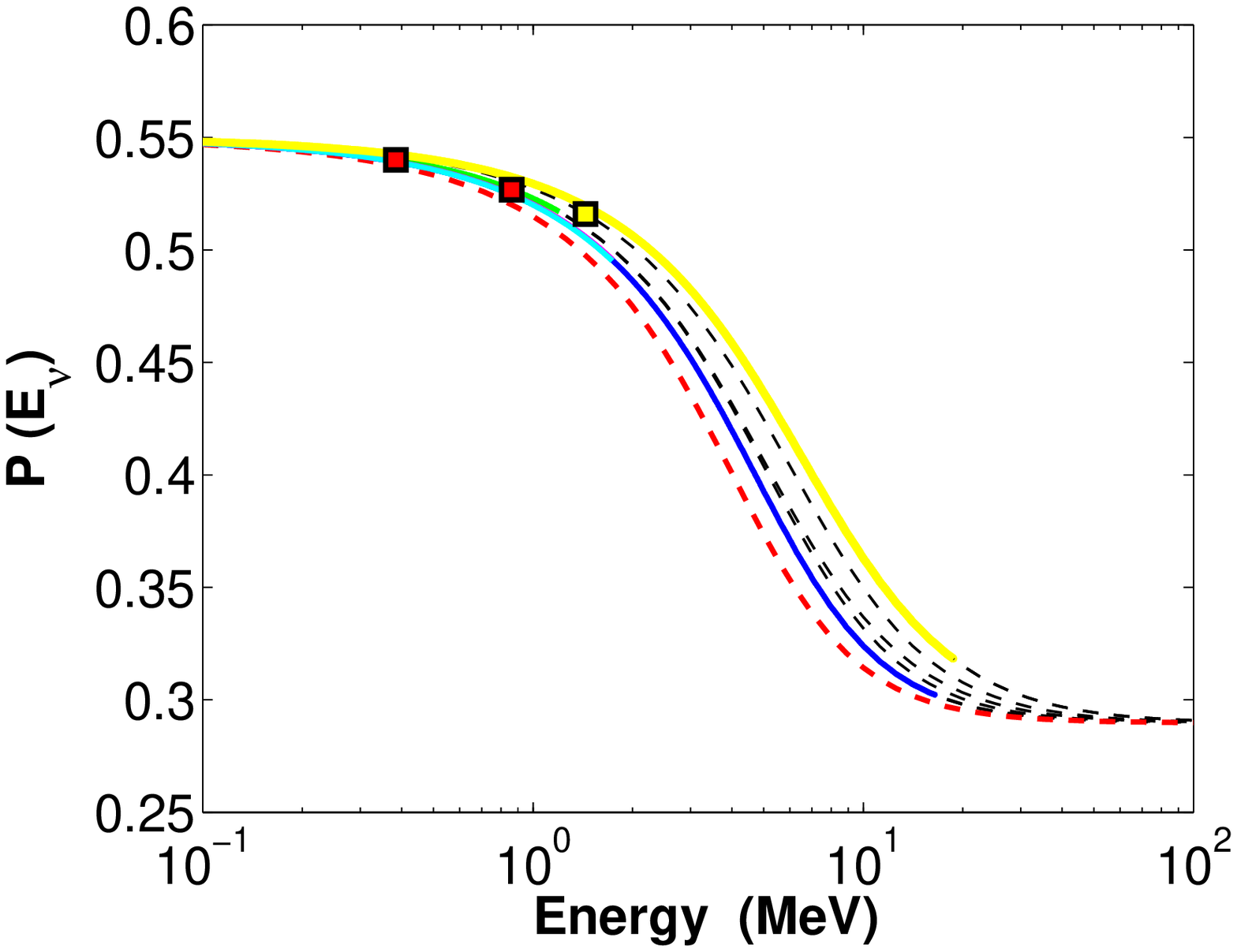}
}
\caption{The survival probability of electron-neutrinos in function of the neutrino energy. 
The reference curve (red dashed curve) defines the survival probability of electron-neutrinos in the centre of the Sun. 
The colored parts of the curves indicate the energy range of neutrinos produced in the Sun's core 
for each nuclear reaction (cf. Fig.~\ref{fig1}).}
\label{fig3}
\end{figure}

Since electron neutrinos are produced in nuclear reactions located at different distances from the center, 
there is a clear differentiation between the different survival probability curves (cf. Fig.~\ref{fig3}).  
Nevertheless, for neutrinos with low or high energy, these curves become indistinguishable, as low-energy neutrinos 
are affected only by vacuum fluctuations, and  high-energy neutrinos are affected by a cumulative effect of vacuum oscillations and matter oscillations. It is worth noticing that the possibility of observing the MSW effect on solar neutrinos is limited, as the neutrino flavor oscillations induced by matter depend on the neutrino emission spectrum (cf. Fig.~\ref{fig1}) and on the location of the neutrino source in the Sun's core (cf. Fig.~\ref{fig2}).
The parts of the survival probability of electron neutrinos that can be measured by solar neutrino experiments are shown in Fig.~\ref{fig3}.
In particular, we notice that the $^8B$ neutrino flavor oscillation 
is strongly dependent on the MSW effect, specially for neutrinos with a higher energy.
However, the $^7Be$ neutrino flux is much less dependent on the MSW effect, 
particularly the $^7Be$ neutrino flux  corresponding to the line of lower energy. 
Similarly, the $pep$ neutrino flux is 
also weakly dependent on the MSW effect.  Accordingly,  
the constraint obtained in the electronic density from the $^8B$ neutrino flux measurement  
is much more reliable than the constraint obtained from  
$^7Be$ and $pep$  neutrino flux measurements (cf. Fig.~\ref{fig3}).
Furthermore, it is worth noticing a second-order effect. The conversion of electron neutrinos due to matter oscillations  depends on the local electron density $n_e(r)$: a decrease of the central density of $n_e(r=0)$ moves
 all the $\langle P_{\nu_e } (E_\nu)\rangle$ of different neutrino sources to the right, and a rapid variation of  $n_e(r)$ (possibly
 caused by a rapid variation of density)  with the radius increases the distance  between the consecutive $\langle P_{\nu_e } (E_\nu)\rangle$ curves (cf. Fig.~\ref{fig3}).

Once   the fundamental parameters of solar neutrino oscillations in vacuum are known and   determined independently from solar neutrino fluxes, in principle,  the probability of survival of electron-flavor solar neutrinos can be used to infer the radial electronic density profile in different locations of the solar core.
By perturbation analysis of Eqs.~(\ref{eq-A}-\ref{eq-F}), the value of the electronic $n$ is  determined for each value of the survival probability of electron neutrinos obtained from the experimental neutrino data~\cite{2013ApJ...765...14L}.
The procedure is as follows: for  neutrinos  of a given energy  $E_\nu$,  the electronic density correction $\Delta n $  
is computed  from the standard solar  model electronic density $n_o$  as  $ \Delta n/ n_o =  \beta_o \; \Delta P/ P_o$, 
with $\Delta P= \bar{P}-P_o$ where $P_o$  and $\bar{P}$ are the survival probabilities of electron  neutrinos obtained
from the  standard solar model and experimental data. $\beta_o$ is a coefficient computed from the  standard solar model data.
It follows that the new value of electronic density, $\bar{n}_e$ is obtained as $\bar{n}_e=n_{o}+\Delta n$.
Figure~\ref{fig4} shows the inverted electronic density values obtained from the  survival probability of electron neutrinos 
computed from experimental data. 

\section{Solar Neutrino Fluxes}

Usually, the  fluxes of the different neutrino flavors are represented by $\Phi_{t}$ for the total neutrino flux,  $\Phi_{e}$  for the electron-neutrino flux, and  $\Phi_{\tau\mu}$  for the nonelectron flavor component of neutrinos, corresponding to the experimentally indistinguishable flavors of $\tau$ and $\mu$ neutrinos~\cite{2010LNP...817.....B}. It follows that $\Phi_{t}=\Phi_{e}+\Phi_{\tau\mu}$. Such quantities are computed from the measured neutrino fluxes that
result from the interaction of neutrinos with the detector, which occur through three different interaction processes. 
As for the charged-current reaction (CC) of neutrinos with a deuterium nucleus and the elastic scattering (ES) of neutrinos off electrons, both processes are observed through the detection of the Cherenkov light produced by electrons in the heavy water. A third process known as neutral-current (NC) reaction is observed via the detection of neutrons. Conveniently, such neutrino fluxes are called $\Phi_{CC}$, $\Phi_{ES}$ and $\Phi_{NC}$.  
$\Phi_{t}$ is determined as $\Phi_{t}=\Phi_{NC}$, once that the neutral-current reaction is sensitive to all
flavors of neutrinos.  The flux of electron-neutrinos is measured by the charged-current reaction  $\Phi_{e}=\Phi_{CC}$, or
alternatively it can  be measured by the neutrino-electron scattering (ES), for which  $\Phi_{e}$ is computed as 
$\Phi_{e}=1.2\Phi_{ES}-0.20\Phi_{NC}$.

Among current solar neutrino experiments, the measurements of $^8B$ neutrino flavors are the most compelling because for $^8B$ neutrinos, unlike in the case of $^7Be$ and $pep$ neutrinos~\cite{2011PhRvL.107n1302B,2008PhRvL.101i1302A,2012PhRvL.108e1302B}, the measurements allow the computation of $\Phi_{t}$ and  $\Phi_{e}$
for different and independent solar experiments. 
Furthermore, $^8B$ neutrino flux is much more sensitive to the density stratification of the solar 
core-- i.e., the
neutrino flavor oscillation induced by the MSW effect, than other sources of neutrino fluxes  
such as $^7Be$ and $pep$ neutrinos.
This allows us to make an estimation of the survival probability of 
electron-neutrino flavors independent of solar models and neutrino oscillation  models (see Table~\ref{tab1}).  
$\Phi_{t}(^8B)$ is estimated by the Sudbury Neutrino Observatory (SNO phase III) experiment~\cite{2013PhRvC..87a5502A} 
from $\Phi_{NC}(^8B)$, a value 7\% larger than the previous SNO measurement~\cite{2010PhRvC..81e5504A}.
The SNO Collaboration have performed another measurement~\cite{2004PhRvL..92r1301A}
of $\Phi_{NC}(^8B)$ by enhancing the sensitivity of heavy water to neutral current interaction (En. NC).

\begin{table}[t]
\caption{$^8B$ neutrino fluxes and electron-neutrino \\ survival probabilities}
\begin{tabular}{l|l|l|l}
\hline
Experiment $[\Phi_{NC}]$   &  $\Phi_{CC}$ or $\Phi_{ES} $ &  $\Phi_{e}$   &   $P_{\nu_e}$    \\
$10^6\; {\rm cm^{-2} s^{-1}}$ & $10^6\; {\rm cm^{-2} s^{-1}}$     &  $10^6\; {\rm cm^{-2} s^{-1}}$   &      \\
\hline
SNO(Phase III) \\  
$5.54\pm 0.69$  \\
\hline
SNO (Phase III)\footnote{Aharmim {\it et al.}~\cite{2013PhRvC..87a5502A}} & $CC$:$1.67\pm 0.12$  & $1.67\pm 0.12$  &  $0.30 \pm 0.04$    \\
SNO (Phase II)\footnote{Aharmim et al. {\it et al.}~\cite{2007PhRvC..75d5502A}} & $CC$:$1.76\pm 0.14$  & $1.76\pm 0.14$  &   $0.32\pm 0.05$    \\
Borexino\footnote{Bellini et al. {\it et al.}~\cite{2010PhRvD..82c3006B}} & $ES$:$2.4\pm 0.5$  & $1.77\pm 0.62$  &   $0.32 \pm 0.12 $    \\
SK (Phase III)\footnote{Abe et al. {\it et al.}~\cite{2011PhRvD..83e2010A}} & $ES$:$2.32\pm 0.09$  & $1.67\pm 0.18$  &   $0.30\pm 0.05$    \\
\hline
SNO (En. NC) \\ 
$5.21\pm 0.65$ \\
\hline
SNO (Phase III) & $CC$:$1.67\pm 0.12$  & $1.67\pm 0.12$  &  $0.32 \pm 0.05$    \\
SNO (Phase II) & $CC$:$1.76\pm 0.14$  & $1.76\pm 0.14$  &   $0.34 \pm 0.05$    \\
Borexino& $ES$:$2.4\pm 0.5$  & $1.84\pm 0.61$  &   $0.35\pm 0.13 $    \\
SK (Phase III) & $ES$:$2.32\pm 0.09$  & $1.74\pm 0.17$  &   $0.33\pm 0.05$    \\
\hline
Mean Value   & $-$  & $- $  &   $0.32\pm 0.20$   \\
\hline
\end{tabular}
\label{tab1}
\end{table}

In Table~\ref{tab1} we show the electron-neutrino flux $\Phi_{e}(^8B)$ estimated from the SNO,
Super-Kamiokande and Borexino measurements~\cite{2007PhRvC..75d5502A,2011PhRvD..83e2010A,2010PhRvD..82c3006B}.
Although the errors in most of the neutrino-measured fluxes are still quite significant,  
it is encouraging to observe that the different experiments  lead to quite identical values of $\Phi_{e}(^8B)$.
The $\Phi_{ES}(^8B)$ computed in the case of SNO phase III predicts a value 4\% lower that in the case of  SNO En. NC.  
The survival probability of electron-neutrinos $P_{\nu_e}(^8B)$ was computed as the ratio $\Phi_{e}(^8B)/\Phi_{t}(^8B)$
\cite{2001PhLB..521..287B}. The error is obtained by adding up the errors quadratically. 
The values of $P_{\nu_e}(^8B)$ are quite similar for all experiments  with a difference smaller than 17\% (Cf. Table~\ref{tab1}).
In particular, the value of the Borexino data  (SNO phase III),  $P_{\nu_e}(^8B)=0.32\pm 0.12$ is 9\% higher 
than the estimation made by the Borexino Team\cite{2010PhRvD..82c3006B}, 
which obtained a value $P_{\nu_e}(^8B)=0.29\pm 0.1$ at the mean energy of 8.9 MeV.
The averaged value of all the experiments $P_{\nu_e}(^8B)$ is estimated to be $0.32\pm 0.20$. 
\begin{figure}
\centering 
\includegraphics[scale=0.5]{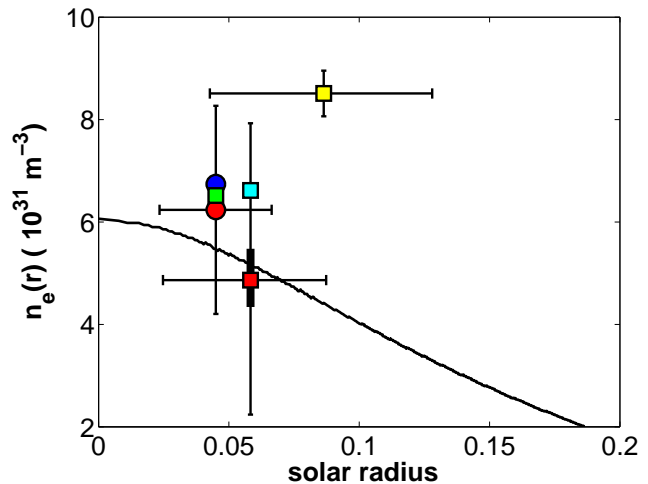}
\caption{
The core of the Sun: the radial profile of electronic density  (black continuous curve).  
The points shown correspond 
to the values  of the electronic density {\it inverted}, $n_e(\bar{r})$, for the values of $\bar{r}$, $0.045\;R_\odot$,  
$0.058\;R_\odot$ and $0.086\;R_\odot$ computed from the survival probability of electron neutrinos for
$^8B$, $^7Be$ and $pep$ neutrino fluxes (see Table~\ref{tab1}):  
{\bf(a)} {\it $^8Be$ neutrino flux} - averaged $P_{\nu_e} (^8B) $ (red sphere), 
 $P_{\nu_e} (^8B)=0.32\pm 0.05 $   (SNO, blue sphere), 
 $P_{\nu_e} (^8B) =0.32\pm 0.12 $  (Borexino, green square).
{\bf (b)}  {\it $^7Be$ and $pep$ neutrino fluxes} (Boroxino experiment), (i) The $P_{\nu_e} (pep) =0.62 \pm 0.17$ (yellow square)~\cite{2012PhRvL.108e1302B}, (ii) The $ P_{\nu_e}(^7Be) =0.52^{+0.07}_{-0.06}$ (red square)  
and $P_{\nu_e}(^7Be) =0.56^{+0.10}_{-0.10}$ (cyan square)~\cite{2011PhRvL.107n1302B,2008PhRvL.101i1302A}. 
The electronic densities inverted from the $^8B$ (red circle) and $pep$ (yellow square) neutrino fluxes were computed assuming that the current vertical errors of $P_{\nu_e} (^8B) $  and $P_{\nu_e} (pep) $ were reduced by a factor 15. In the case of the $^7Be$ (red square) neutrino flux, this was computed using the real vertical error bar -- the thicker bar here corresponds to a reduction of the current error by a factor of 5.}
\label{fig4}
\end{figure}  
Figure~\ref{fig4} shows the {\it inverted} electronic density  $\bar{n}_e$ $(\equiv n_e(\bar{r}))$ at 
different locations of the solar radius, computed as  described in the previous section.  
The theoretical value $P_{o}$ was estimated  
for  neutrinos with energy above 5 MeV in the case of  $^8B$ neutrinos, 
and for the values of $0.862$ MeV and  $1.44$ MeV for $^7Be$ and $pep$ neutrinos.
In Fig~\ref{fig4}, for reasons of clarity, the error bars are shown only in three cases. 
In all these data points, the length of the  {\it horizontal  error bar}  of each inverted data point $\bar{n}_e$ 
(of a specific $\bar{r}$) defines the radial interval where 68.2\% (equivalent to one $\sigma$, in a normal distribution) 
of the neutrino flux is produced.

The mean value of $P_{\nu_e}(^8B)$ and most of the individual values 
suggest that the electronic density in the core of the Sun is at least 
25\% higher than in the current solar model (cf. Table~\ref{tab1} and Fig.~\ref{fig4}).
The high value of $P_{\nu_e}(pep)$ obtained by the Borexino Team reinforces the high value 
of $n_e(r)$ in the core. Although the present value of $P_{\nu_e}(^7Be)$
is in agreement with the  standard solar model,  the previous determination 
$P_{\nu_e}(^7Be)$ also suggests a high value of $n_e(r)$ in the core.  Nevertheless, it should be noticed that 
both electronic density values  obtained from $pep$ and $^7Be$ survival electron-neutrino probabilities,
depend on the solar model (contrarily to $^8B$), and therefore have a limited diagnostic capability. 

A potential confirmation of our findings can be achieved as the experimental error in the neutrino measurements decreases and levels of accuracy significantly increase. As per Fig.~\ref{fig4}, a clear insight will be possible if the error in the determination of  the electron-neutrino survival probability  obtained from the present measurements can be reduced by a factor  of  15.  

\section{Summary and Conclusion}

We show   that if the fundamental parameters of neutrino oscillations are determined from  Earth's neutrino detector  flux  measurements,  as  per the  KamLAND  experiment, then  the solar neutrino fluxes can be used to invert the electronic density in the  Sun's inner core.  In this work we have developed  a new method to infer the electronic density  based in this principle.
The method consists in determining the {\it real} electronic density of the solar core (in function of the radius) 
as a small correction to the electronic density predicted by the standard solar model.

All the observed neutrino fluxes, i.e.,   the $^8B$, $^7Be$ and $pep$  neutrino fluxes have neutrino emission sources located within 10\% of the solar radius. Although the accuracy of  the  current neutrino flux measurements  is low,  we have found that  a significant improvement  in the accuracy of these  measurements will allow  the  determination of the electronic density
in  the  Sun's inner core with an error smaller than a few percent.  
In particular,  the reduction of the error bar of  the  $^7Be$ electron-neutrino  survival probability by a factor 5, allows the determination of the electronic density with an error of 3\% (cf. Fig.~\ref{fig4}). 
The measurements  of $hep$,  $^{13}N$, $^{15}O$ and  $^{17}F$  neutrino  fluxes 
will also contribute significantly to improve the quality of inversion of the electronic density in the Sun's core. 
In particular  $^8B$ and $hep$ high-energy neutrino fluxes are very
sensitive to matter oscillations (cf. Fig.~\ref{fig3}).  Therefore accurate measurements  of these neutrino fluxes will  also  put stringent constraints to  the  electronic density in the Sun's core.

This diagnostic of the electronic density of the Sun's inner core combined with the accurate 
determination of the abundances 
of heavy elements such as carbon, nitrogen and oxygen from the neutrino fluxes produced in the CNO cycle 
provides presently the best way to probe the physics of the Sun's inner core. This possibility 
will become feasible  with the upgrade of experiments such as Borexino and the Sudbury Neutrino Observatory (SNO+)~\cite[][]{2011NuPhS.217...50M},
as  well as the new  solar neutrino detector Low Energy Neutrino  Astrophysics (LENA)~\cite[][]{2012APh....35..685W}.

\begin{acknowledgments}
This work was supported by grants from  Funda\c c\~ao para 
a Ci\^encia e Tecnologia   and   Funda\c c\~ao Calouste Gulbenkian.
\end{acknowledgments}


\end{document}